\documentclass[iop,numberedappendix]{emulateapj}

\usepackage{amsmath}
\usepackage{mathrsfs}
\usepackage{empheq}
\usepackage{url}
\usepackage{multirow}
\usepackage{dcolumn}   
\usepackage{amssymb}   
\usepackage{hyperref}
\usepackage[applemac]{inputenc}
\usepackage{graphicx,subfigure}
\usepackage{bm}
\usepackage{color}

\shorttitle{Tearing modes in partially ionized astrophysical plasma}
\shortauthors{Fulvia Pucci, K.A.P. Singh, Marco Velli, Anna Tenerani}

\begin{document}

\title{Tearing modes in partially ionized astrophysical plasma}

\author{Fulvia Pucci\altaffilmark{1,2}, K. Alkendra P. Singh \altaffilmark{3,4}, Anna Tenerani \altaffilmark{5}, Marco Velli \altaffilmark{6}} 
\email{fulvia.pucci@lasp.colorado.edu}
\email{alkendra.solarastrophysics@gmail.com, singh@kwasan.kyoto-u.ac.jp}
\affil{\altaffilmark{1} LASP, University of Colorado Boulder,1234 Innovation Dr, Boulder, CO 80303}
\affil{\altaffilmark{2} National Institute for Fusion Science, National Institutes of Natural Sciences,
Toki 509-5292, Japan}
\affil{\altaffilmark{3} Department of Physics, Institute of Science, BHU, Varanasi 221005, India}
\affil{\altaffilmark{4} Astronomical Observatory, Graduate School of Science, Kyoto University , Yamashina, Kyoto 607-8471, Japan}
\affil{\altaffilmark{5} Department of Physics, University of Texas, Austin, TX 78712, USA}
\affil{\altaffilmark{6}Department of Earth, Planetary, and Space Sciences, UCLA,Los Angeles, 595 Charles E Young Dr E, Los Angeles, CA 90095}

\begin{abstract}
In many astrophysical environments the plasma is only partially ionized, and therefore the interaction of charged and neutral particles may alter both the triggering of reconnection and its subsequent dynamical evolution.
We derive the tearing mode maximum growth rate for partially ionized plasmas in the cases of weak and strong coupling between the plasma and the neutrals. In addition, critical scalings for current sheet aspect ratios are presented in terms of Lundquist number and ion-neutral collision frequencies. In the decoupled regime the standard tearing mode is recovered with a small correction depending on the ion-neutral collision frequency; in the intermediate regime collisions with neutrals are shown to stabilize current sheets, resulting in larger critical aspect ratios for ideal tearing to occur. Nonetheless, the additional electron-neutral collisions, hidden in the definition of the Lundquist number, can shrink the critical aspect ratios below the fully ionized case. In the coupled regime, the growth rate depends on the density ratio between ions and neutrals through the collision frequency between these two species.  These provide critical aspect ratios for which the tearing mode instability transitions from slow to ideal, that depend on the neutral-ion density ratio. 
\end{abstract}

\section{Introduction}
Magnetic reconnection is considered to be an important dynamical mechanism in a variety of  astrophysical plasmas \citep{ZweibelYamada:2009, Yamadaetal:2010}. Without magnetic reconnection, stars and accretion disks would not have coronae, magnetic dynamos would not work, and there would most probably be no supersonic solar wind e.g. \citep{ZweibelYamada:2009, Yamadaetal:2010}. A complete understanding of magnetic reconnection in astrophysical settings therefore requires explaining how energy accumulates in the magnetic field, how current carrying fields becomes unstable, and how magnetic energy release occurs on short timescales once the reconnection process has been triggered. 
One of the major difficulties in understanding magnetic reconnection in astrophysical plasmas stems from the fact that classical models of reconnection, starting from the steady state Sweet-Parker mechanism \citep{parker:1957, Sweet:1958}, or the non-steady, resistive instabilities \citep{FKR:1963}, appeared to be inadequate to explain the observed, transient and explosive release of magnetic energy. 
More recently, the thin Sweet-Parker current sheets have been shown to be unstable to a fast tearing instability \citep{Biskamp:1986, ShibataTanuma:2001,Loureiro:2007}. \citet{PucciVelli:2014} (hereafter PV14) showed that, in a resistive framework, current sheet inverse aspect ratios scaling as $a/L \sim S^{-1/3}$,  separates slowly evolving systems from ones that are so unstable they should never form. In the expression $a$ is the thickness and $L$ is the length of the current sheet, $S =L V_A /\eta_m$, with $\eta_m$ the magnetic diffusivity, is the Lundquist number. They called this regime Ideal Tearing (IT hereafter). This scaling was confirmed in numerical simulations by \citet{Landietal:2015,Teneranietal:2015b,Landi17, Huangetal:2017}, and extended to recursive reconnection and general plasmoid number scalings by \citet{Singhetal:2019}.\\ Additional effects, such as kinetic effects, could be also become important \citep{Singhetal:2015,DelSartoetal:2016,Puccietal:2017}.
There are also environments (e.g. solar photosphere and solar chromosphere, solar filaments/prominences, the interstellar medium, dense molecular clouds, protoplanetary discs) where the astrophysical plasmas undergoing reconnection are only partially ionized (see e.g. \citet{Ballester:2018}).
The ionization degree depends upon the electron-neutral and the electron-ion collision frequencies (\citet{AlfvenFalthammer:1962}), while the resulting drag force acting on each species must satisfy momentum conservation for the whole plasma. This means that depending on the ratio between the density of the ions and neutrals (or electrons and neutrals, if their collisions are not negligible), the associated collision frequencies may establish additional characteristic times scales of the system. Depending on the strength of the coupling between ion and neutrals and the dynamical times of magnetic reconnection, the reconnection rate can be affected.
There are a number of theoretical studies on tearing mode instability that show the dependence of the growth rate of the instability on ion-neutral collisions (\citet{Zweibel:1989,Zweibeletal:2011,Singhetal:2015}).\\
Multi-fluid MHD simulations show that, as a result of current sheet thinning and elongation, a critical Lundquist number $S_{c}$, is reached in a partially ionized plasma, at which point plasmoid formation starts (\citet{Leakeetal:2012,Leakeetal:2013}).
In such multi-fluid simulations, during the current sheet thinning, a stage is reached where the neutrals and ions decouple, and a reconnection rate faster than the single-fluid Sweet-Parker prediction is observed. The ion and neutral outflows are well coupled in the multi-fluid MHD simulations in the sense that the difference between ion and neutral outflow is negligible compared to the magnitude of the ion outflow.
Assuming incompressibility and the same pressure gradient for ion and neutrals, in a reduced MHD frame, \citet{Zweibel:1989} calculated the growth rate of the classic tearing instability in the so called \textit{constant-psi} regime \citep{FKR:1963}.
In this paper, starting from the model described in \citet{Zweibel:1989}, we calculate the maximum growth rate for the tearing mode instability in partially ionized plasmas, assuming as a primary source of drag the collisions between ions and neutrals (retaining Coulomb collisions between ions and electrons). We  calculate the scaling of the growth rate depending on the coupling, the relative speed of collisions and the growth rate itself. Then, applying the IT criterion, we find, for each regime, the scaling of the critical aspect ratio for which the growth rate depends neither on the Lundquist number, nor the density ratios. 


\section{Tearing modes in a partially ionized plasma}
%
%
Consider a one-dimensional current sheet structure in which the magnetic field reverses sign
\begin{equation}
\label{harris}
\vec B(y) = B(y)\hat{i}= B_0 F \left( \dfrac{y}{a}\right)\hat{i},
\end{equation}
where $B_{0}$ is the asymptotic amplitude of the field, $F$ is an arbitrary odd non-dimensional function, whose first derivative provides the current profile. A specific example is given by the Harris current sheet $ F={\tanh}(y/a)$. The dispersion relation for the reconnecting tearing instability depends, in the resistive magnetohydrodynamics (MHD) framework, on the magnetic diffusivity $\eta$, the shear-scale $a$ defining the current sheet thickness, the wavenumber $ka$.
As discussed in \citet{DelSartoetal:2016, Puccietal:2018} for general equilibrium profiles, speci\-fying the function $F$ results in a different dependence on the wavenumber $ka$. This arises from the fact that at
%
large Lundquist number two regions define the solution structure: a boundary layer of thickness 2$\delta$ around the center ($y = 0$) of the current sheet, and outer regions where diffusivity and growth rate may be neglected. Such outer solutions
%
lead to a discontinuity of the first derivative of the perturbing magnetic field at the neutral point (regularized by diffusion in the inner layer): the jump in the gradient of the reconnecting field component is 
called $\Delta'$. Two asymptotic expressions summarize the dispersion relation, depending on whether $\Delta' \delta/a \ll 1$ (small Delta prime or $\Delta'$, subscript SD), where 
\begin{equation}\label{eq:SD}
 \gamma_{_{SD}}\bar{\tau}_A\simeq A^{\frac{4}{5}}\bar{k}^{\frac{2}{5}}(\Delta')^{\frac{4}{5}}\bar{S}^{-\frac{3}{5}}\qquad {\delta_{_{SD}}/a}\sim
(\bar{S}\bar{k})^{-\frac{2}{5}}(\Delta')^{\frac{1}{5}},
\end{equation}
where $A$ is a non-dimensional constant, or
$\Delta' \delta/a \gg 1$ (large Delta prime or $\Delta'$, subscript LD)
\begin{equation}
\label{eq:LD}
\gamma_{_{LD}}\bar{\tau}_A\simeq \bar{k}^{\frac{2}{3}}\bar{S}^{-\frac{1}{3}}\qquad {\delta_{_{LD}}/ a}\sim (\bar{S}\bar{k})^{-\frac{1}{3}},
\end{equation}
in which case the growth rate no longer depends explicitly on $\Delta'$ \citep{DelSartoetal:2016}. Here, barred quantities are normalized to the current sheet \emph{thickness} ($\bar \tau_A = a/V_A$, $\bar k = ka$, $\bar S = aV_A/\eta_m$). The expressions above may be used to find the scaling of the fastest growing mode by assuming that both relations remain valid at the wave-number of maximum growth $k_m(\bar S)$ for sufficiently large $\bar S$. 
For the Harris current sheet for which $\Delta' \sim 2/ka$ this implies 
\begin{equation}
\label{maxtear}
\gamma \bar \tau_A \sim \bar S^{ -\frac{1}{2}}\,,\,\,\,\frac{\delta}{a} \sim \bar S^{ -\frac{1}{4}}\,,\,\,\,
k_m a\sim \bar S^{-\frac{1}{4}}.
\end{equation}
The relation for the ``ideal'' tearing or (IT) instability, i.e. for an instability where the growth rate survives independently of the Lundquist number in the ideal limit \citep{PucciVelli:2014}, is obtained by rescaling the dispersion relation to the current sheet length rather than the thickness
\begin{equation}
\label{harrisdisp}
\gamma \tau_A \sim S^{-\frac{1}{2}}\left({\frac{a}{L}}\right)^{-\frac{3}{2}}.
\end{equation}
Assuming an inverse aspect ratio of the form $a/L\sim S^{-\alpha}$, any value of $\alpha<1/3$ leads to a divergence of growth rates in the ideal limit, while any value of $\alpha>1/3$ leads to growth rates which tend to zero as the Lundquist number grows without bound \citep{PucciVelli:2014}. This result is very general: any additional effect, such as viscosity \citep{Teneranietal:2015}, Hall current \citep{Puccietal:2017}, will result in a different critical aspect ratio scaling at which fast reconnection is triggered.

\subsection{Modifications due to ion-neutral interactions}
In a partially-ionized plasma, the effect of electron-neutral and the electron-ion collisions on the plasma dynamics is the generation of an Ohmic type diffusion. In the presence of three different species undergoing collisions, the single fluid description may apply in the partially ionized limit, with an appropriately modified magnetic induction equation.
\\
Considering three different species (electrons, ions and neutrals) the momentum conservation for each of the three species may be written separately, including inter-species collision terms, neglecting ionization and recombination effects. In \citet{Zweibel:1989} the Coulomb collisions between ions and electrons reflect in an Ohmic diffusion coefficient in the induction equation that remains the same as in the fully ionized case. We notice here that, as shown in \citet{SinghKrishan:2010} the actual value of the resistivity is enhanced if the electron neutral collisions are taken into account, but the Ohmic resistivity is substantially calculated in the same way, yielding a magnetic diffusivity:
\begin{equation}
\label{etaohm}
\eta_{m}=\dfrac{c^2}{\omega^2_{pe}}(\nu_{ei}+\nu_{en})
\end{equation}
where $\omega_{pe}$ is the electron plasma frequency, the electron ion and electron neutral collision frequencies are $\nu_{ei,en}$ and $c$ is the speed of light. In \citet{Zweibel:1989} the interaction of the plasma with neutrals occurs through ion-neutral collisions, while electron neutral collisions are not taken into account. In this way, the tearing equation for the momentum conservation of ion and neutrals combined writes (primes denote derivatives with respect 
to the $a$-scaled variable $y/a$):
\begin{eqnarray}
\label{piptear}
(\gamma \bar \tau_{Ai})^2\left( 1+\dfrac{\nu_{in}}{\gamma+\nu_{ni}}\right)(\phi''- \bar k^2\phi)=\nonumber\\
-F(\psi''-\bar k^2\psi)+F''\psi
\\
\psi=\bar kF\phi+ \frac{1}{\bar S\gamma \bar \tau_{Ai}} (\psi''-\bar k^2\psi),\nonumber
\end{eqnarray}
and $\bar \tau_{Ai}$ is the Alfv\'en time calculated with the ion density (still normalized to the sheet thickness $a$), $\gamma$ is the tearing growth rate associated with a mode with wave vector $\bar k=ka$ along the equilibrium magnetic field. The collision frequencies are calculated assuming binary elastic (energy and momentum conservation) collisions between electrons and neutrals so that $\nu_{ni} = \dfrac{n_im_i}{n_nm_n} \ \nu_{in} \Rightarrow \nu_{ni} < \nu_{in}$ at most heights in the solar atmosphere (see Tab. 1 in \citet{Singhetal:2015}). Note that the opposite limit $\nu_{ni} \gg \nu_{in}$ leads to the standard tearing of a completely ionized plasma.
Following \citet{Zweibel:1989} we may redefine a starred Alfv\'en time and Lundquist number 
\begin{eqnarray}
\label{Zweibelpres}
\bar \tau_{Ai} (1+\frac{\nu_{in}}{\gamma +
\nu_{ni}})^{1/2}&:=& \bar \tau_{Ai} f_M^{1/2}   \rightarrow \bar{\tau}_A^{*},\\
\bar S^*&:=& \bar S \frac{\bar \tau_{Ai}}{\bar{\tau}_A^{*}}.
\end{eqnarray}
Note that when the growth rate is negligible compared to both collision frequencies, the factor $f_M^{1/2}$ becomes
\begin{equation}
f_M^{1/2}=(1+\frac{\nu_{in}}{\nu_{ni}})^{1/2}=(1+\frac{\rho_{n}}
{\rho_{i}})^{1/2}=(\frac{\rho}{\rho_{i}})^{1/2},
\end{equation}
where $\rho=\rho_i+\rho_n$ is the total mass density. In other words, in this limit $\bar{\tau}_A^{*}=\bar{\tau}_A$, the 
Alfv\'en time based on the Alfv\'en speed $V_{AT}$ calculated with the total (ion plus neutral) mass density, and the Lundquist number $\bar S^*$ also reduces to the Lundquist number based on the Alfv\'en speed calculated with the total density.

In \citet{Zweibel:1989} the modified tearing mode analysis is carried out only in the small $\Delta'$ regime, see eqs. \eqref{eq:SD}. Here we analyze the tearing mode equations considering the maximum growth rate of the tearing instability eq.\eqref{maxtear}, because the fastest growing mode is the most relevant in the context of triggering fast magnetic reconnection in natural plasmas. Inserting $\bar{\tau}_A^{*}$ into eqs. \eqref{piptear}, the latter regain the exact standard form for the tearing mode with the substitution of $\bar S \bar \tau_{Ai}$ with $\bar S^*\bar{\tau}_A^{*}$ and $\gamma\bar \tau_{Ai} $ with $\gamma \bar{\tau}^*_A$ , so that from eq.\eqref{maxtear} we have that $\gamma \bar{\tau}_A^{*}$ follows the same scaling with $\bar S^*$ as in the standard tearing theory:
\begin{eqnarray}
\label{standardgrowth}
\gamma \bar{\tau}^*_A \propto (\bar S^*)^{-1/2} \Rightarrow \gamma \bar \tau_{Ai}\propto(\bar S)^{-1/2}\left(\dfrac{\bar \tau_{Ai}}{\bar{\tau}^*_A}\right)^{1/2}
\end{eqnarray}
Following Zweibel 1989; Singh et al. 2019 we can define three different regimes, ordered by the magnitude of the growth rate relative to the neutral-ion and ion-neutral collision frequencies:
\\

\noindent {\bf 1. Coupled regime} $\gamma \ll \nu_{ni} \ll \nu_{in}$: in this regime, eq.(\ref{standardgrowth}) simply means that the fastest tearing mode growth rate, normalized to the total density-based Alfv\'en time,
scales in the standard way with the total density-based Lundquist number, as we commented above.
The result may be written
\begin{equation} 
\label{coupled3} 
\gamma \bar \tau_{Ai} \sim \bar S^{-1/2} \left(\dfrac{\rho}{\rho_{i}}\right)^{-1/4}. 
\end{equation}
Table 1 in \citet{Singhetal:2015} shows that the ratio $\rho_{n}/\rho_{i}$ can be up to $10^6$ in some of the solar atmospheric layers. For such cases of interest the dispersion relation becomes $\gamma \bar \tau_{Ai} \sim \bar S^{-1/2} \left(\dfrac{\rho_{n}}{\rho_{i}}\right)^{-1/4}$.
\\

\noindent {\bf 2. Intermediate regime} $\nu_{ni} \ll \gamma \ll \nu_{in}$: ion neutral collisions partly couple the ionized and neutral fluids, introducing an effective, stabilizing viscous drag on the plasma so that the growth rate is reduced, and\\
$\gamma \bar \tau_{Ai} \sim \bar S^{-1/2}\left(\dfrac{\nu_{in}}{\gamma }\right)^{-1/4} =S^{-1/2}\left(\dfrac{\nu_{in}\bar \tau_{Ai}}{\gamma \bar \tau_{Ai}}\right)^{-1/4},$ implying $ \gamma \bar \tau_{Ai} \sim \bar S^{-2/3}\left(\nu_{in}\bar \tau_{Ai}\right)^{-1/3} $.
The reconnection rate depends on the ratio between the ion-neutral collision time and the Alfv\'en time.
\\

\noindent{\bf 3. Uncoupled regime} $\nu_{ni} \ll \nu_{in}\ll \gamma$: the current sheet is thin enough that the fastest growing tearing mode on the ionic component is too fast for ions and neutrals to communicate collisionally, so at lowest order $\gamma \bar{\tau}_A^{*} \sim \gamma \bar \tau_{Ai}$. Corrections of order $\nu_{in}/\gamma$ can be found. From eq.\eqref{Zweibelpres}, eq. \eqref{standardgrowth} becomes:
 \begin{eqnarray}
\label{growthUNCOUPLED}
&&\hspace{0.7cm} \gamma \bar \tau_{Ai} \propto\bar S^{-1/2}\left(1+\dfrac{1}{2} \dfrac{\nu_{in}}{\gamma}\right)^{-1/2}\hspace{-0.5cm}\sim \bar S^{-1/2}\left(1-\dfrac{1}{4} \dfrac{\nu_{in}}{\gamma}\right)\nonumber
\end{eqnarray}
where $\dfrac{\nu_{in}}{\gamma}:=\epsilon\ll 1$ and we neglected terms of order $\epsilon^2$, leading to
\begin{equation}
\label{growthCOUPLED2}
\gamma \bar \tau_{Ai}\propto \bar S^{-1/2} - \dfrac{1}{4} \nu_{in}\bar \tau_{Ai} .
\end{equation}

The discussion above is written in terms of a growth rate that is a function of the plasma parameters, and so it is not immediately clear how such parameters determine the appropriate instability regime. This is clarified by analyzing the 3 regime inequalities in a bit more detail.

The timescale condition of the coupled regimes $\gamma \ll \nu_{ni}$ may be rendered explicitly using the growth rate calculated in Eq.(\ref{coupled3}), so that 
$$\bar S^{-1/2}\left(\dfrac{\rho}{\rho_{i}}\right)^{-1/4}\ll \nu_{ni}\bar{\tau}_{Ai}. $$ By writing out
the explicit form of the Lundquist number and timescales, this
translates directly into a condition on the current sheet thickness:
$$
a\gg a_{c1} = \left({\eta_m V_{AT}}/{\nu_{ni}^2}\right)^{1/3}.
$$ The critical thickness $a_{c1}$ exists because of the new intrinsic timescale given by the neutral-ion collision time. 

The timescale inequality of the intermediate regime 
$\gamma \ll \nu_{in}$ now translates into a second condition on the current sheet thickness $a\gg a_{c2} = \left({\eta_m V_{Ai}}/{\nu_{in}^2}\right)^{1/3}$.

The ratio of the two critical thicknesses is
$a_{c1}/a_{c2}= (\rho_n/\rho_i)^{1/2}$. In other words, in order of decreasing current sheet thickness, one transitions from the coupled into the intermediate regime. If the current sheet is much thinner than $a_{c2}$ one ends up in the decoupled regime, where the tearing modes grows on the ionized component timescales and Lundquist numbers, with the neutrals not contributing in any way.%

In the next subsection we describe the initiation of reconnection within a framework of a dynamics driven by the corresponding fast or Alfv\'enic timescales, making use of the results just obtained. 
\subsection{The Ideal Tearing mode in partial ionized plasmas.}
Following \citet{PucciVelli:2014}, we renormalize the Alfv\'en time and Lundquist number using a macroscopic length scale of the system $L$ in place of the equilibrium magnetic field scale $a$, i.e. $S:=Lv_{Ai}/\eta$ and $\tau_{Ai}=L/v_{Ai}$. The same normalization holds for the Lundquist and Alfv\'en times based on the total (ion plus neutral density).

The three regimes discussed above may now lead to different critical aspect ratios for the initiation of a Lundquist-number \emph{and} neutral to ionic density) independent growth rate.
\\

\noindent {\bf 1. Coupled regime:} The ``ideal'' tearing (IT) criterion is applied to the growth rate and Lundquist number, both normalized to the Alfv\'en time based on the total density. All quantities are rescaled from the thickness $a$ to the (macroscopic) length $L$. 

Quantitatively:
\begin{eqnarray}
\label{final2}
&{S}^{2}(\rho_n/\rho_i)^{-1} :=P_D, \dfrac{a}{L}\sim (P_D)^{-\zeta}\\
&\Rightarrow  \ (P_D)^{-1/4}(P_D)^{3/2\zeta}\sim O(1)
\end{eqnarray} 
yielding a critical aspect ratio scaling as
\begin{eqnarray}
\label{threshold2}
\dfrac{a_c}{L}\sim (P_D)^{-1/6}&=({S}^{2}(\rho_i/\rho_n))^{-1/6}=\\
&{S}^{-1/3}(\rho_n/\rho_i)^{1/6},
\nonumber
\end{eqnarray}
where the subscript \textit{c} indicates the critical current sheet thickness (which must still be compared with the intrinsic scales $a_{c1,c2}$). For the solar atmosphere the density dependence means the inverse aspect ratio can be up to  10 times targer than the fully ionized IT critical inverse aspect ratio \citep{Singhetal:2015}.
This regime is attained only if  
$a_c\gg a_{c1}$.
\\

\noindent{\bf 2. Intermediate regime:} the IT criterion leads to $\gamma \tau_{Ai} \sim O(1)$
\begin{eqnarray}
\label{harris1}
\gamma \tau_{Ai} \sim{S}^{-2/3}
\left(\dfrac{a}{L}\right)^{-6/3}(\nu_{in}\tau_{Ai})^{-1/3}\sim O(1)
\end{eqnarray}
Defining
\begin{eqnarray}
\label{final}
&{S}^{2}(\nu_{in}{\tau}_{Ai}) :=P_p, \dfrac{a}{L}\sim (P_p)^{-\zeta}\\
&\Rightarrow  \ (P_p)^{-1/3}(P_p)^{6/3\zeta}\sim O(1) \\
\end{eqnarray}
we obtain $\zeta = 1/6$ meaning
\begin{eqnarray}
\label{threshold}
\dfrac{a_c}{L}\sim ({S}^{2}(\nu_{in}{\tau}_{Ai}))^{-1/6}= {S}^{-1/3}(\nu_{in}{\tau}_{Ai})^{-1/6}
\end{eqnarray}
The dependence on the Lundquist number is the same as the classical IT. The additional factor gives a smaller critical inverse aspect ratio scaling than in the fully ionized case.  One may think of this as an effective viscosity due to the "sloshing" between ions and neutrals on these timescales. Since we are in the intermediate regime $\nu_{in}\gg \gamma\gg \nu_{ni} \Rightarrow  \nu_{in} {\tau}_{Ai} \gg 1$. This shouldn't actually come as a surprise, considering $\bar{\tau}_A^{*} > \tau_{Ai}$, see eq.\eqref{Zweibelpres}.
Still, as described by eq. \eqref{etaohm}, if the electron-neutral collisions are high enough to significantly lower the Lundquist number, the critical aspect ratio could be actually higher, i.e. the presence of the neutrals destabilizes larger sheets than the fully ionized ones. Finally, note that the previously derived intermediate regime inequality on the critical aspect ratio thickness must still hold ${a_c}\gg a_{cr2}$.
\\

\noindent{\bf 3. Uncoupled regime:} 
This regime holds when the current sheet is very thin, and as we have seen above this means that $a_c$ is much smaller than $a_{c2}$. The corrections to the standard IT tearing criterion depend only weakly on the small values of $\nu_{in}\tau_{Ai}$. The IT assumption  $\gamma \tau_{Ai} \sim O(1)$ so, in this regime fast reconnection is triggered with the neutrals not really noticing. 

\subsection{The inner resistive layer.}
The region around the neutral sheet, where the perturbations to the background field are significant, is the inner resistive layer $\delta$ (see, e.g. \citet{Puccietal:2018}). This parameter is particularly important because when $\delta$ becomes of the order of the kinetic scales, kinetic effects play a role in the reconnection dynamics (see e.g. \citet{Terasawa:1983,Puccietal:2017}).  In \citet{Zweibel:1989} an estimation of $\delta$ is given and the dependence on the ion-neutral collision frequency is recovered. In our case the expression for the maximum growth rate is given in eq. \eqref{maxtear}, where in the partial ionized case $\delta / a \sim \bar{S}^{*-1/4}=(\tau_D/\bar{\tau}_A^{*})^{-1/4}=\bar{S}^{-1/4} f_M^{-1/8}$. Since $f_M$ is invariant for the IT rescaling, the solution is the same as for the classic IT with corrections depending on the regime, $\delta / L =S^{-1/2} f_M^{-1/8}$. In particular, for the solar atmosphere, in the decoupled regime, the additional factor  \citep{SinghKrishan:2010} tells us the inner resistive layer is slightly larger than in the fully ionized case so that kinetic effects are less likely to kick in. 
%
\section{Summary and Conclusion}
In this paper we have discussed the onset of fast reconnection in partially ionized plasmas, considering three species undergoing collisions: ions, electrons and neutrals. The ionization degree depends on the relative collision frequencies and we neglected the effect of ionization and recombination. Assuming as in \citet{Zweibel:1989} the interaction with neutrals occurs through the ion-neutral collisions, we considered the combined ion and neutral equation of motion and the magnetic induction equation as the system describing the Tearing instability of a generic equilibrium configuration. The magnetic diffusivity is also implicitly modified due to the additional collisions between neutrals and electrons. Since we wanted to study the onset of fast magnetic reconnection we derived the scalings for the Tearing maximum growth rate for three different regimes: coupled, intermediate and decoupled. We calculated the inverse aspect ratio, for which the growth rate does not depend on the collision rates (and so the Lundquist number). \\

In the coupled regime, the critical aspect ratio depends on the ratio between the neutral density and the ion density. The dependence is weak, however, since $\rho_n/\rho_i$ may be as large as $10^6$ in the solar corona \citep{SinghKrishan:2010}, the critical current sheet thickness can be up to 10 times larger than in the fully ionized case. \\

In the intermediate regime, the scaling with the Lundquist number remains the same as in the fully ionized case. A dependence on $\nu_{in}\tau_{Ai}$ arises, which causes the critical inverse aspect ratio to scale with a correction that makes it appear to be smaller than in the fully ionized case. However, the intrinsic thickness of the sheet
remains thicker than in the decoupled regime,
as shown by the inequalities between $a_{c}$, $a_{c1}$ and $a_{c2}$. In addition, the role of electron-neutral collisions may significantly change the Lundquist number, resulting in a destabilizing effect.\\

Finally, in the decoupled regime a small correction ($\propto \nu_{in}\tau_A$) arises with respect to the fully ionized case. This results in small corrections (factor $<10$) to the critical aspect ratio.\\

\section{Acknowledgements}
We would like to thank Prof. Kazunari Shibata for fundamental discussions and insights on the physics and trigger of magnetic reconnection.
KAPS gratefully acknowledges the UGC Faculty Recharge Program of Ministry of Human Resource Development (MHRD), Govt. of India and University Grants Commission (UGC), New Delhi as well as the visiting associateship program of Inter University Centre for Astronomy \& Astrophysics (IUCAA), Pune.
AH is supported by his STFC Ernest Rutherford Fellowship grant number ST/L00397X/2 and by STFC grant ST/R000891/1. M.V. was supported by the NSF-DOE Partnership in Basic Plasma Science and Engineering award N.1619611 and the NASA Parker Solar Probe Observatory Scientist grant NNX15AF34G.
This research was supported in part by the National Science Foundation under Grant No.NSF PHY-1748958.

\bibliography{Bib_PIP}
\bibliographystyle{apalike}

\end{document}